\documentclass[11pt, draftclsnofoot, onecolumn]{IEEEtran}  %

\usepackage{subfigure}
\usepackage{pdfpages}
\usepackage{diagbox}
\usepackage{colortbl}
\definecolor{mygray}{gray}{0.85}
\usepackage{array}
\usepackage{caption}
\usepackage[linesnumbered,boxed,ruled,commentsnumbered]{algorithm2e}

 \usepackage{eqparbox}

\usepackage{amsmath,bm}
\usepackage{cite}
\usepackage{amsmath,amssymb,amsfonts}
\usepackage{algorithmic}
\usepackage{amsthm}
\usepackage{graphicx}
\usepackage{textcomp}
\usepackage{xcolor}

\usepackage{hyperref}
\hypersetup{%
  bookmarks=true
}

\begin{document}

\title{\huge
Intelligent Spectrum Learning for Wireless Networks with  Reconfigurable Intelligent Surfaces
}

\author{
\IEEEauthorblockN{Bo~Yang,
 Xuelin Cao, Chongwen Huang, Chau Yuen,~\IEEEmembership{Fellow,~IEEE}, \\ Lijun Qian,~\IEEEmembership{Senior Member,~IEEE},  and Marco Di Renzo, \IEEEmembership{Fellow,~IEEE}
 }
\thanks{This paper has been accepted by IEEE TVT.}
 \thanks{B. Yang, X. Cao, and C. Yuen are with the Engineering Product Development Pillar, Singapore University of Technology and Design, Singapore 487372 (e-mail: bo$\_$yang, xuelin$\_$cao, yuenchau@sutd.edu.sg).}%
 \thanks{C. Huang is with Zhejiang Provincial Key Lab of information processing, communication and networking, Zhejiang University, No.38 Zheda Road, Hangzhou, 310007, P.R. China (e-mail: chongwenhuang@zju.edu.cn).}
  \thanks{L. Qian is with the Department of Electrical and Computer Engineering and CREDIT Center, Prairie View A$\&$M University, Texas A$\&$M University System, Prairie View, TX 77446, USA (e-mail: liqian@pvamu.edu).} 
\thanks{M. Di Renzo is with Universit\'e Paris-Saclay, CNRS, CentraleSup\'elec, Laboratoire des Signaux et Syst\`emes, 3 Rue Joliot-Curie, 91192 Gif-sur-Yvette, France. (marco.di-renzo@universite-paris-saclay.fr)}

 }

\maketitle

\begin{abstract}
Reconfigurable intelligent surface (RIS) has become a promising technology for enhancing the reliability of wireless communications, which is capable of reflecting the desired signals through appropriate phase shifts. However, the intended signals that impinge upon an RIS are often mixed with interfering signals, which are usually dynamic and unknown. In particular, the received signal-to-interference-plus-noise ratio (SINR) may be degraded by the signals reflected from the RISs that originate from non-intended users. To tackle this issue, we introduce the concept of intelligent spectrum learning (ISL), which uses an appropriately trained convolutional
neural network (CNN) at the RIS controller to help the RISs infer the interfering signals directly from the incident signals. By capitalizing on the ISL, a distributed control algorithm is proposed to maximize the received SINR by dynamically configuring the active/inactive binary status of the RIS elements. Simulation results validate the performance improvement offered by deep learning and demonstrate the superiority of the proposed ISL-aided approach.
\end{abstract}



\section{Introduction}
Due to the dynamic nature of the wireless environment that results in severe signal fluctuations caused by multipath fading and the presence of large obstacles, the wireless link between a desired cellular user and a base station (BS) may not be reliable enough or may even undergo a complete outage. To tackle this issue, reconfigurable intelligent surfaces (RISs) have been proposed to improve the received signal-to-interference-plus-noise ratio (SINR) at the users by appropriately reflecting the incident signals and generating directional beams~\cite{EURASIP}.

In the literature, some preliminary works investigated the optimization of RIS-assisted wireless communications. In~\cite{Huang02}, a joint transmit power allocation and phase shift design was developed to maximize the energy efficiency. 
In~\cite{WU01}, the authors considered a downlink RIS-assisted multiuser communication system and studied a joint transmission and reflection beamforming problem to minimize the total transmit power. \textcolor{black}{In~\cite{letter_add}, the channel estimation problem was investigated for an RIS-aided wireless communication system by jointly optimizing the training sequence of the transmitter and the reflection pattern of the RIS.} Furthermore, an RIS-assisted anti-jamming solution was proposed for securing wireless communications via reinforcement learning~\cite{globecom}. In~\cite{jsac}, the authors introduced RISs in mobile edge computing systems, where a joint design of computing and communications was developed to minimize the computational latency.

\begin{figure}[t]
  \captionsetup{font={footnotesize }}
\centerline{ \includegraphics[width=5.27in, height=3.25in]{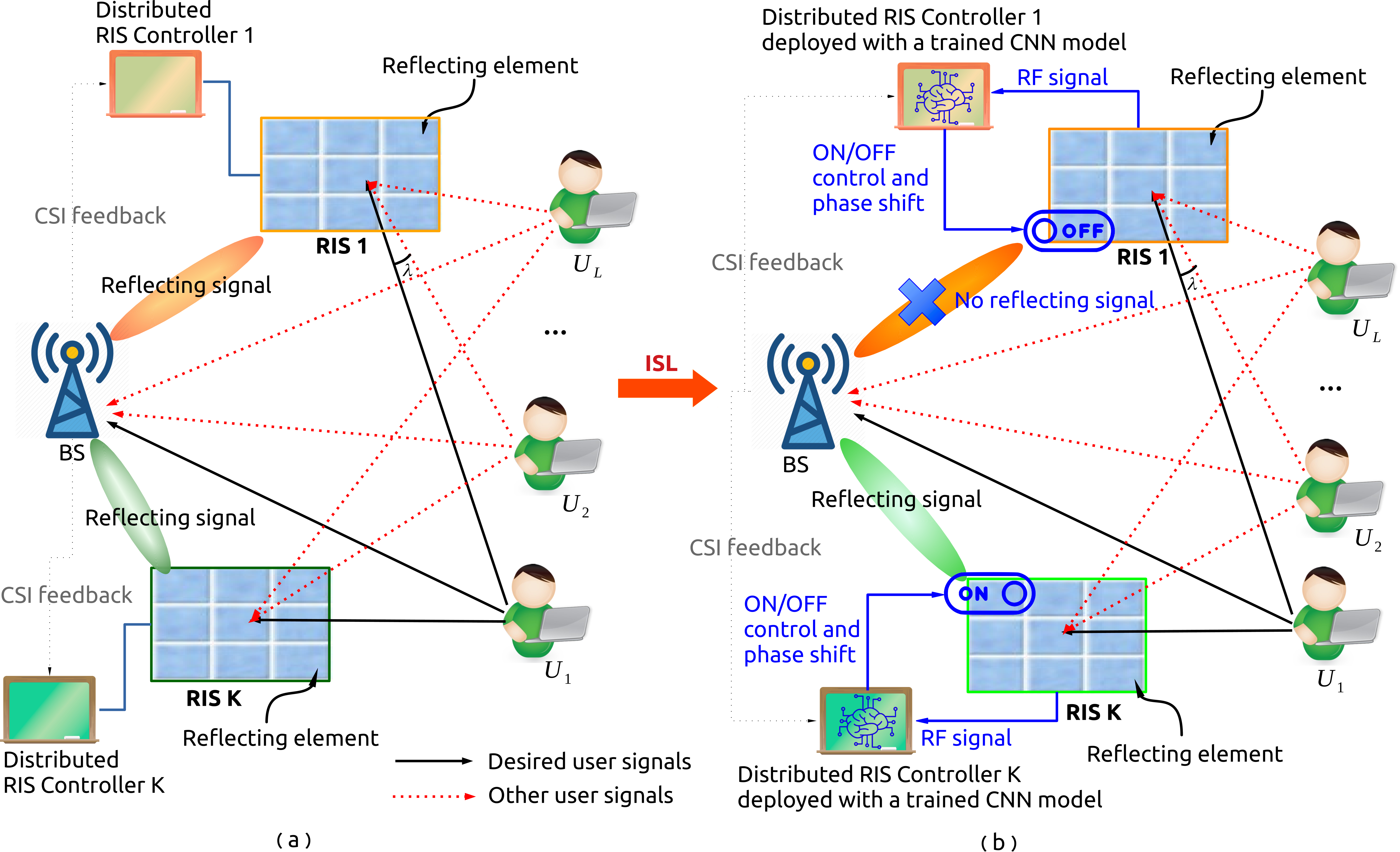}}
\caption{In (a), a traditional multiple-user uplink RIS-assisted wireless communication system with $K$ RISs is shown, where we assume that $L$ users transmit to the BS at same time and frequency, and all RISs serve one user at a time. So, the other users are considered as interferers. In (b), the proposed ISL-enabled RIS-assisted wireless communication system is shown. Each RIS controller is deployed with a trained CNN model to identify the interfering users from the incident RF signals, so as to optimize the operation of the RISs in a distributed way. 
}
\vspace{-5.8mm}
\label{motivation_case}
\end{figure}

However, most of the existing works assume either that no interference exists, which rarely occurs in practice, or that the interference is known and can be taken into account, which is not trivial to estimate since the interference is usually dynamically changing. These issues are exacerbated in RIS-aided systems, since they are nearly-passive surfaces with no active sensing capabilities for channel and interference estimation~\cite{Renzo}. 
In order to elucidate the problem at hand, let us consider the case study depicted in Fig.~\ref{motivation_case}(a), in which we consider a practical scenario where all the RISs serve one \textit{desired user }(e.g., $U_1$). In the analyzed case study, the other users (e.g., $U_l$, $l\in[2,L]$) are considered as \textit{interfering users} for $U_1$ at each RIS. 
As a result, the signal reflected by each RIS is a mixture of the desired signal from $U_1$ and the interfering signals from the other users. {In this context, the received SINR at the BS via an RIS that does not account for the interfering signals could be even worse than the SINR of the direct link.} This impact of the interference is, in particular, more severe if the interfering devices are close to the desired user, e.g., in Fig.~\ref{motivation_case}(a) $U_2$ may cause severe interference to $U_1$ at RIS$_1$ when the angle between them (i.e., $\lambda$) is small.

To overcome these challenges, in this paper, we empower a conventional RIS-assisted wireless system with `intelligent spectrum learning (ISL) capabilities', by leveraging appropriately trained convolutional neural networks (CNN) at the RIS controller in order to predict/estimate the interfering devices from the incident signals, as highlighted in Fig.~\ref{motivation_case}(b). In the proposed system, the active-inactive (or ON-OFF) status of each RIS\footnote{In this paper, the ON-OFF status is referred to having the entire RIS ON or OFF, i.e., either all the elements of the RIS are turned ON or all the elements of the RIS are turned OFF.} and the corresponding phase shifts need to be carefully optimized, since the interference distribution at each RIS is, in general, different. The corresponding SINR maximization problem turns out to be a mixed-integer nonlinear program (MINLP), which is usually difficult to solve. To tackle this issue, we decompose the original problem into two subproblems, which are solved independently and in a distributed manner.
The proposed solution equips a conventional RIS with the capability to dynamically `{think-and-decide}' whether reflecting or not  the incident signals through the proposed distributed ISL principle.


\section{System Model and Problem Formulation} \label{system_model}
We consider an RIS-assisted uplink wireless system that consists of one BS, a set $\cal K$ of $K$ RISs, and a set $\cal L$ of $L$ users, where $K \geq 1$ and $L>K$ usually hold. We assume that the BS allocates all the RISs to serve one user at a time in order to improve the quality of the wireless link. Also, the direct links between the users and the BS are available. 
Under these assumptions, all the other users act as interferers for the intended user either through the direct links or through the links reflected by the RISs. Each RIS operates as a nearly-passive surface, i.e., the RIS elements are passive but the RIS controller may consume power~\cite{Huang01}. In particular, the RIS controller is equipped with a CNN, as shown in Fig.~\ref{motivation_case}(b).  The CNN is discussed in the next sections.
\vspace{-3mm}
\subsection{RIS-Assisted Communication Model}
Each RIS, denoted as R$_k$, $k \in {\cal K}$, is equipped with $N_k$ reflecting elements, which can be appropriately configured by the RIS controller to reflect the signals of the users towards the BS. In general, the RISs can be appropriately deployed so that line-of-sight (LoS) links can be established with the BS and, possibly, the users. \textcolor{black}{We assume that the channel state information (CSI) of all channels involved is perfectly known at the BS\footnote{\textcolor{black}{Many papers in the literature have tackled the issue of estimating and reporting the CSI, e.g.,~\cite{Renzo,CSI}. Therefore, this problem is not addressed in this paper and it is left to a future research work.}}, which, in turn, can feed back the CSI to the RIS controller via a dedicated control channel~\cite{WU01,jsac}.}  


Each RIS can be in two possible states: ON and OFF. We introduce a binary variable $\beta_k \in \{0,1\}$ to indicate the ON-OFF state of R$_k$. $\beta_k = 1$ indicates that R$_k$ is ON, which means that it reflects the incident signals, while $\beta_k = 0$ indicates R$_k$ is OFF, which means that it does not reflect any signals. 

As for the $k$th RIS that is ON, the amplitude reflection coefficient is assumed to be equal to one for all the $N_k$ reflecting elements and the phase reflection matrix is 
\begin{equation} \label{phi}
\mathbf{\Phi}_k={\rm{diag}} \left (e^{j \theta_1^k}, e^{j \theta_2^k},...,e^{j \theta_{N_k}^k} \right ),
\end{equation} 
where $\Theta_k=\left (\theta_1^k, \theta_2^k,..., \theta_{N_k}^k \right )$ denotes the vector of phase shifts that can be optimized by R$_k$.

\vspace{-4mm}
\subsection{Wireless Channel Model}
We consider an RIS-aided uplink wireless system, where the channels from the desired user (denoted as $U_l$) to R$_k$, from R$_k$ to the BS, and from the $m$th interfering user (denoted as $U_m$, $m \in {\cal L}, m \neq l$) to R$_k$, are $\textit{\textbf h}_{l,k} \in \mathbb{C}^{N_k \times 1}$, $\textit{\textbf g}_{k} \in \mathbb{C}^{1 \times N_k}$, and $\textit{\textbf h}_{m,k} \in \mathbb{C}^{N_k \times 1}$, respectively. The channel gains of the direct links from $U_l$ to BS and from $U_m$ to BS are denoted by $h_{d,l}$ and $h_{d,m}$, respectively. These channels are assumed to be perfectly estimated and quasi-static, hence remaining nearly-constant during the transmission time~\cite{WU01}. 

Without loss of generality, we assume that the users are randomly distributed and have time-varying traffic demands. This implies that the users may not be all active during the considered transmission time\footnote{We assume that the active interfering users remain unchanged during the transmission time of the desired user.}. In particular, the total number of interfering users for $U_l$ is given by 
$\omega_l = \sum_{m \in {\cal L}, m \neq l} \alpha_m$,
 where $\alpha_m \in \{0,1\}$ is a binary variable that indicates that $U_m$ is active ($\alpha_m=1$) or inactive ($\alpha_m=0$) and therefore can cause or not interference to $U_l$, respectively. 
  
 Accordingly, the set of active interfering users is ${\cal W}\!=\!\{U_m|\alpha_m\!=\!1\}$, where $\forall m \neq l$ and $\forall m \in {\cal L}$.   
 Since the number of active interferers is dynamic, it is not easy to estimate the interference distribution over time. The received signal at the BS depends on the desired signal sent from
$U_l$ (including the direct link and the link reflected by the RIS), the interference from the interfering users in ${\cal W}$, and the white Gaussian noise:
\begin{equation} \label{recv_sig}
\begin{array}{*{20}{l}}
y_l = \underbrace{\left(h_{d,l}+ \sum_{k=1}^{K} \beta_k \textit{\textbf g}_{k} \mathbf{\Phi}_k \textit{\textbf h}_{l,k} \right) \sqrt{p_l} s_l}_{{\rm Desired \ signal \ from} \ U_l} + 
\underbrace{\sum_{m \in {\cal W}} \left(h_{d,m}\!+\! \sum_{k=1}^{K} \xi_{m,l}^k \beta_k \textit{\textbf g}_{k} \mathbf{\Phi}_k \textit{\textbf h}_{m,k} \right) \sqrt{p_m} s_m}_{\rm Interference \ from \ other \ users} + n_l,
\end{array}
\end{equation}
where $p_l$ and $p_m$ denote the transmit power of $U_l$ and $U_m$, respectively, $s_l$ and $s_m$ are the unit-power information signals sent from $U_l$ and $U_m$, respectively, and $n_l \sim {\cal C N }(0, \sigma^2)$ is the white Gaussian noise. In addition, $\xi_{m,l}^k \in [0,1]$ accounts for the impact of the interference caused by $U_m$ to $U_l$ and that is associated to the $k$th RIS, as clarified in \textbf{Assumption~\ref{R0}}.

\theoremstyle{Remark}
\theoremstyle{assumption}
\newtheorem{remark}{{Remark}}
\newtheorem{assumption}{{Assumption}}
\begin{assumption} \label{R0} 
We assume that the impact of the strength of the interference reflected by an RIS is {inversely} proportional to the difference of the angles of incidence between the desired user and the interfering users at the RIS (see $\lambda$ in Fig. 1(b) for an example). 
\end{assumption}
\vspace{-5mm}

\subsection{Problem Formulation and Analysis}
Let ${\cal B}\!=\!\{\beta_1, \beta_2,..., \beta_K\}$ denote the RIS binary decision vector that collects the binary variables that identify the ON-OFF status of the RISs.
The set of RISs that are active is ${\cal R}\!=\!\left \{R_k | \beta_{k}\!=\!1\right \}, \ \forall k \in {\cal K}$.
 Accordingly, the SINR at the BS for the intended user $U_l$ is 
\begin{equation} \label{sinr}
\gamma_l\!=\! \frac{{p_l} \left |h_{d,l}+ \sum_{k=1}^{K} \beta_k \textit{\textbf g}_{k} \mathbf{\Phi}_k \textit{\textbf h}_{l,k}  \right |^2 }{\sum_{m=1}^{\omega_l}{p_m} \left |h_{d,m} \!+\! \sum_{k=1}^{K} \xi_{m,l}^k \beta_k \textit{\textbf g}_{k} \mathbf{\Phi}_k \textit{\textbf h}_{m,k}  \right |^2 \!+\! \sigma^2}.
\end{equation}

Our objective is to maximize the SINR in (\ref{sinr}), by optimizing the RIS binary activation vector (${\cal B}$), and the phase shifts matrix of the active RISs which is denoted as ${\bf  \Theta} = \{\Theta_1, \Theta_2, ..., \Theta_K\}$ . To this end, we need to solve the following optimization problem:  
\vspace{-3mm}
\begin{subequations} \label{problem}
\begin{align}
 P_1: & \;\;\;\;\underset{{\cal B}, {\bf \Theta}}{\rm max}\;\; \gamma_l  \\
& \;\;\;\;\;\;{\rm{s}}{\rm{.t}}{\rm{.}}\;\; \beta_{k}\in \left \{0,1  \right \}, \  \forall k \in {\cal K}, \\
& \;\;\;\;\;\;\;\;\;\;\;\;\;  \left |e^{j \theta_{n}^i} \right | = 1, \ \forall n \in [1, N_i], \ \forall i \in {\cal R}.
 \end{align}
\end{subequations}

Constraint (\ref{problem}b) indicates that the $k$th RIS can only be ON (i.e., $\beta_{k}=1$) or OFF (i.e., $\beta_{k}=0$) at one time. Constraint (\ref{problem}c) indicates that each RIS reflecting element can only provide a phase shift $\theta_{n}^i \in [0, 2\pi)$ without amplifying the signals. 

\textbf{Our Proposal:} 
We observe that $P_1$ is an MINLP, which is NP-hard and whose global optimal solution is, in general, difficult to obtain. In addition, traditional optimization methods may be computationally intensive. 
An emerging approach to tackle this issue is to apply {deep learning} methods to solve $P_1$ at a reduced computational complexity~\cite{{YB_TMC},{LY}}. $P_1$, however, may not be easy to solve even using deep learning methods, since the number of interfering users is a random variable that is unknown. \textcolor{black}{In addition, during the channel estimation phase, the RISs cannot estimate on their own the active interferers because they  only reflect the incident signals in a passive manner.} This makes the solution of $P_1$ even more difficult. To address this challenge, we propose a distributed control mechanism that solves $P_1$ with the aid of the ISL algorithm.

\vspace{-2.0mm}
\section{Distributed RIS Control via Intelligent Spectrum Learning}
\label{IIS}
In this section, the ISL-aided distributed RIS control mechanism is introduced. The proposed approach leverages appropriately trained CNNs at the controller of the RISs, which can identify the active interfering users in a distributed way. 
\vspace{-3.0mm}

\subsection{Intelligent Spectrum Learning}                                                                                                                                                                                                                                                                                                                                                                                                                                                                                                                                                                                                                                                                                                                                                                                                                                                                                                                                                                                                                                                                                                                                                                                                                                                                                                                                                                                                                                                                                                                                                                                                                                                                                                                                                                                                                                                                                                                                                                                                                                                                                                                                                                                                                                                                                                                                                                                                                                                                                                                                                                                                                                                                                                                                                                                                                                                                                                                                                                                                                                                                                                                                                                                                                                                                                                                                                                                                                                                                                                                                                                                                                                                                                                             
ISL is a multi-class classification algorithm that, based on a CNN, returns the set of interfering users for each intended user $U_l$. To design the ISL algorithm, three main aspects need to be discussed: 1) RF traces collection, 2) offline CNN training, and 3) online CNN inference.

\begin{figure}[t]
\centering
  \captionsetup{font={footnotesize }}
\subfigure[]{
\includegraphics[width=2.95in, height=1.1in]{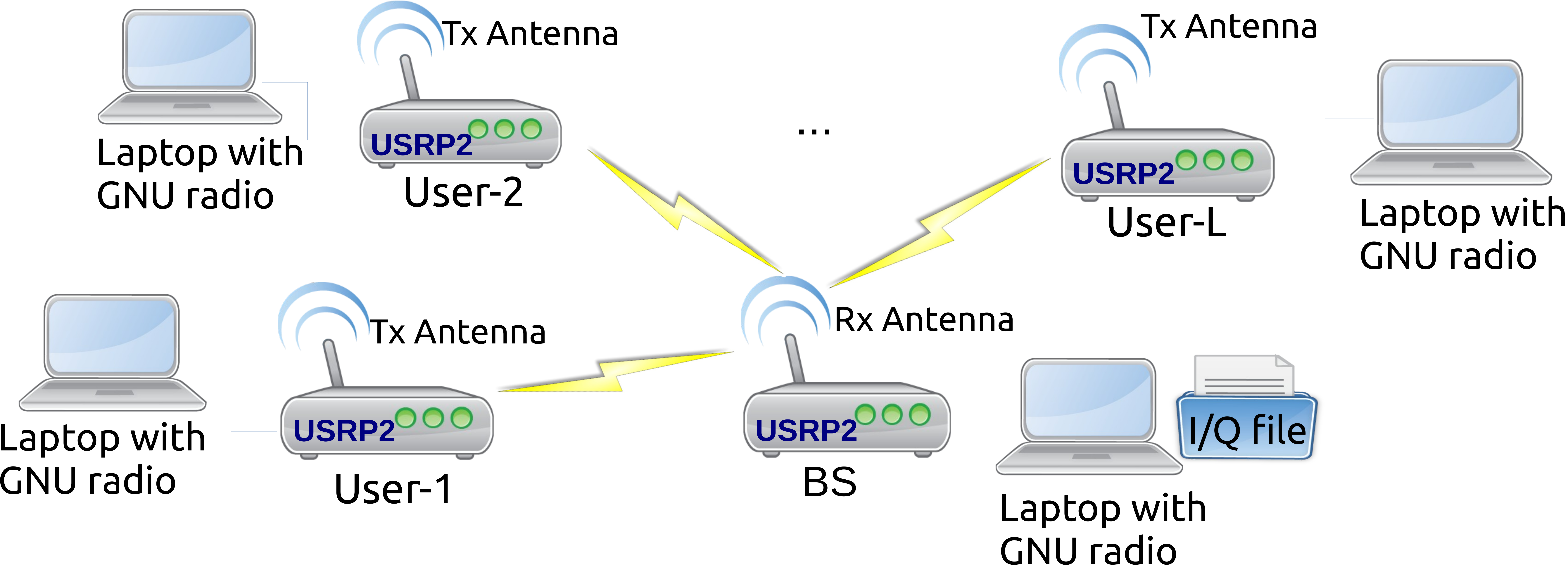}}
\hspace{2mm}
\subfigure[]{
\includegraphics[width=3.2in, height=1.1in]{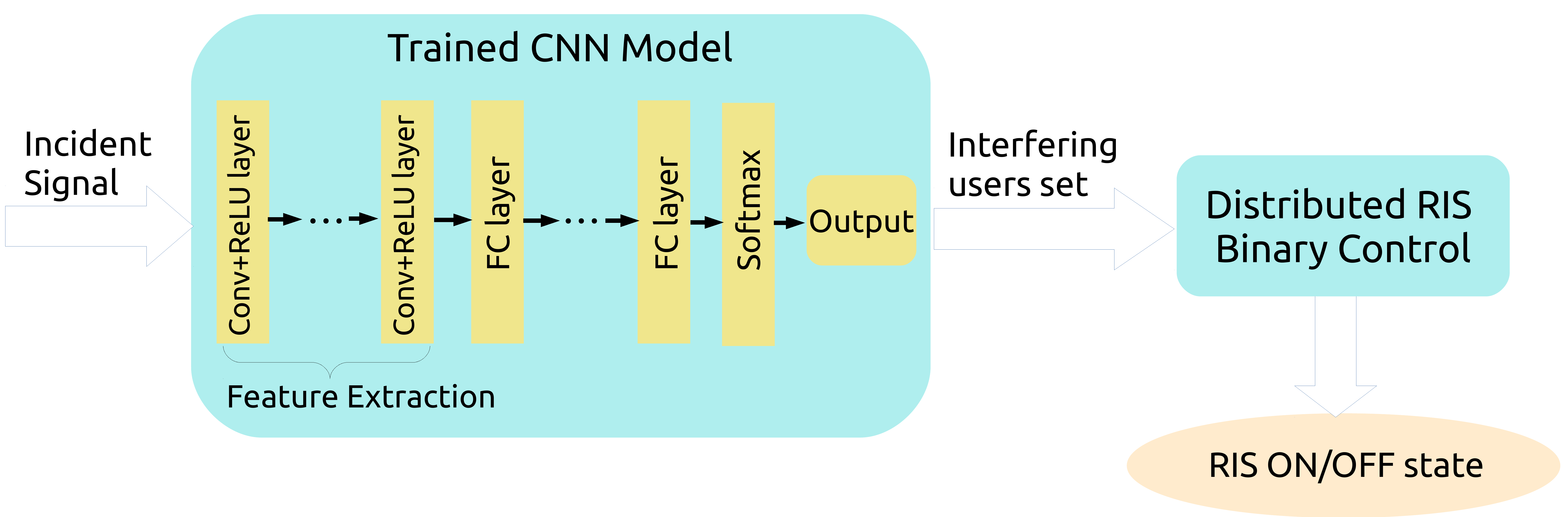}}
\vspace{-0.1in}
\caption{RF trace collection scenario is shown in (a), where $L$ transceivers (users) are scheduled for transmission towards a receiver (BS). The proposed ISL-aided RIS control structure is shown in (b), where the input to the CNN is the incident RF signals and the output is the set of interfering users.} \vspace{-0.5cm}
\label{cnn}
\end{figure}

\vspace{-1.5mm}
 \subsubsection{RF Traces Collection} As far as the RF data collection phase is concerned, historical RF traces are collected using a universal software radio peripheral (USRP2) testbed, \textcolor{black}{which is wired connected (e.g., Gigabit Ethernet) to a host PC with an implementation of the GNU Radio, as illustrated in Fig.~\ref{cnn}(a). In particular, the users are emulated through a laptop that is mainly responsible for baseband processing while a USRP2 platform is used for the up-conversion, the digital-to-analog (D/A) conversion, and wireless transmission of the signals. As far as the BS is concerned, another USRP2 module first receives the signals from the radio interface and then performs A/D and down-conversion. Subsequently, the laptop receives the signals from the USRP2 via the Ethernet and executes the baseband processing. Finally, the Inphase (I) and Quadrature (Q) sequences are stored as a file. In particular, the experimental setup for RF data collection using the USPR2 is performed by using signals at $2.4$ GHz carrier frequency with $1$ MHz bandwidth. In order to collect realistic RF signals in the presence of interference, we let multiple USRP2 units transmit RF signals to an USRP2 receiver. The RF traces have been collected as I/Q sequences,} by including the wireless channel, for a wide range of SNR (e.g., from $0$ to $20$ dB with interval of $5$ dB) in order to account for different interfering cases~\cite{TWC_YB}.
                                                                                                                                                                                                                                                                                                                                                                                                                                                                                                                                                                                                                                                                                                                                                                                                                                                                                                                                                                                                                                                                                                                                                                                                                                                                                                                                                                                                                                                                                                                                                                                                                                                                                                                                                                                                                                                                                                                                                                                                                                                                                                                                                                                                                                                                                                                                                                                                                                                                                                                                                                                                                                                                                                                                                                                                                                                                                                                                                                                                                                                                                                                                                                                                                                                                                                                                                                                                                                                                                                                                                                                                                                                                                                                
 \subsubsection{CNN Offline Training}                                                                                                                                                                                                                                                                                                                                                                                                                                                                                                                                                                                                                                                                                                                                                                                                                                                                                                                                                                                                                                                                                                                                                                                                                                                                                                                                                                                                                                                                                                                                                                                                                                                                                                                                                                                                                                                                                                                                                                                                                                                                                                                                                                                                                                                                                                                                                                                                                                                                                                                                                                                                                                                                                                                                                                                                                                                                                                                                                                                                                                                                                                                                                                                                                                                                                                                                                                                                                                                                                                                                                                                                                                                                                                                  The acquired RF traces have been used for training the CNN architecture illustrated in Fig.~\ref{cnn}(b), where each convolutional layer is followed by a rectified linear units activation function for feature extraction. Fully-connected (FC) layers are used to classify the signals by using the {softmax} activation function for the output layer~\cite{TCOM_Renzo}. The training algorithms is based on the Adam algorithm that uses the cross entropy as the loss function. The CNN model is trained offline using TensorFlow on a GPU cluster (NVIDIA Tesla P100-PCIE-16GB).
                                                                                                                                                                                                                                                                                                                                                                                                                                                                                                                                                                                                                                                                                                                                                                                                                                                                                                                                                                                                                                                                                                                                                                                                                                                                                                                                                                                                                                                                                                                                                                                                                                                                                                                                                                                                                                                                                                                                                                                                                                                                                                                                                                                                                                                                                                                                                                                                                                                                                                                                                                                                                                                                                                                                                                                                                                                                                                                                                                                                                                                                                                                                                                                                                                                                                                                                                                                                                                                                                                                                                                                                                                                                                                            
                                                                                                                                                                                                                                                                                                                                                                                                                                                                                                                                                                                                                                                                                                                                                                                                                                                                                                                                                                                                                                                                                                                                                                                                                                                                                                                                                                                                                                                                                                                                                                                                                                                                                                                                                                                                                                                                                                                                                                                                                                                                                                                                                                                                                                                                                                                                                                                                                                                                                                                                                                                                                                                                                                                                                                                                                                                                                                                                                                                                                                                                                                                                                                                                                                                                                                                                                                                                                                                                                                                                                                                                                                                                                                            \textcolor{black}{Even though the training of the CNN does not account for all possible channel conditions, the generalization property of deep learning enables the trained CNN to infer channel conditions not included in the training dataset~\cite{pac}. 
                                                                                                                                                                                                                                                                                                                                                                                                                                                                                                                                                                                                                                                                                                                                                                                                                                                                                                                                                                                                                                                                                                                                                                                                                                                                                                                                                                                                                                                                                                                                                                                                                                                                                                                                                                                                                                                                                                                                                                                                                                                                                                                                                                                                                                                                                                                                                                                                                                                                                                                                                                                                                                                                                                                                                                                                                                                                                                                                                                                                                                                                                                                                                                                                                                                                                                                                                                                                                                                                                                                                                                                                                                                                                                             It is also noteworthy that several methods have been proposed to scale up the training process of deep neural networks across GPU clusters, which helps to further reduce the runtime of the offline training. Once the CNN model is appropriately trained, it can directly infer incident signals in near real-time. In other words, the proposed ISL-based framework moves the complexity from online computation to offline training.}

\subsubsection{CNN Online Inference}
The CNN is, in particular, trained in order to return the set of active interfering users based on different input signals at each RIS. \textcolor{black}{Specifically, the received RF signals first undergo A/D conversion and frequency down-conversion. Then the baseband I/Q sequences are fed into the trained CNN model to perform online inference at the RIS controller.} 
 
By performing feed-forward calculation via the CNN model (i.e., online inference), the intefering users set for the desired user $U_l$ is obtained as 
\begin{equation} \label{if_set} 
\widetilde{\cal I}_l = \begin{cases}
\{U_m|\widetilde{\alpha}_m=1\}, \ \forall m \in {\cal L}, \forall m \neq l,  & \text{ If } \widetilde{\omega}_l \geq 1, \\ 
\varnothing , & \text{ If } \widetilde{\omega}_l=0,
\end{cases}
\end{equation}
 where $\widetilde{\alpha}_m$ denotes the inferred state flag of $U_m$, and $\widetilde{\omega}_l$ indicates the inferred total number of intefering users in the incident signal.

\subsubsection{An Illustrative ISL Example}
To better understand the proposed ISL algorithm, we illustrate an example with only two users (denoted as $U_1$ and $U_2$), 
\textcolor{black}{and each user has a binary state, i.e., `active'/`ON' and `inactive'/`OFF'. In this case, there exist four combinations of signals from the perspective of each RIS: (1) `Idle' (indicating that both $U_1$ and $U_2$ are inactive), (2) `Only \textit{$U_1$}' (indicating that only $U_1$ is active), (3) `Only \textit{$U_2$}' (indicating that only $U_2$ is active), and (4) `$U_1\!+\!U_2$' (indicating that both $U_1$ and $U_2$ are active). Based on the superimposed incident signal(s), the RIS needs to identify the composition of the signal(s), i.e., to identify the correct class out of the four possible classes of signals. Therefore, this signal identification boils down to a four-class classification problem},  as illustrated in Table~\ref{IIS_result}.


\begin{table}[tbh]  
\center
  \captionsetup{font={small}} 
\caption{An illustrative ISL example with two users} 
\label{IIS_result}  
\begin{tabular}{|l | l|}  
\hline
\textbf{Inferred Class} & \textbf{Description}\\
\hline 
Class-1: Idle & The collected RF traces include only the noise \\
\hline
Class-2: Only $U_1$ & The collected RF traces include only $U_1$ \\
\hline
Class-3: Only $U_2$ & The collected RF traces include only $U_2$ \\
\hline
Class-4: $U_1\!+\!U_2$ & The collected RF traces include both $U_1$ and $U_2$\\
\hline
\end{tabular}  
\end{table} 
\vspace{-5.0mm}

\subsection{Distributed RIS Binary Control}
By feeding the inferred set of interfering users into the distributed RIS binary control algorithm, the corresponding phase shifts and the binary ON-OFF status of the RISs can be obtained, as illustrated in Fig.~\ref{cnn}.

\subsubsection{Optimal Phase Shifts Calculation}
To calculate the phase shifts at the RIS, 
we denote the obtained CSI associated to the $k$th RIS as $\mathbf{C}_k \!=\! \{\textit{\textbf h}_{l,k}, \textit{\textbf h}_{m,k}, \textit{\textbf g}_{k}, h_{d,l}, h_{d,m}\}$. 
Based on the inferred interfering users (including $\widetilde{\cal I}_l$ and $\widetilde{\omega}_l$) obtained via the trained CNN, the received SINR of the signal sent from $U_l$ is given by
\begin{equation} \label{sinr_k}
\widetilde{\gamma}_l \!=\! \frac{{p_l} \left |h_{d,l}+ \sum_{k=1}^{K} \beta_k \textit{\textbf g}_{k} \mathbf{\Phi}_k \textit{\textbf h}_{l,k}  \right |^2 }{\sum_{m=1}^{\widetilde{\omega}_l}{p_m} \left |h_{d,m} \!+\! \sum_{k=1}^{K} \xi_{m,l}^k \beta_k \textit{\textbf g}_{k} \mathbf{\Phi}_k \textit{\textbf h}_{m,k}  \right |^2 \!+\! \sigma^2}.
\end{equation}

Based on (\ref{sinr_k}), we first obtain the phase shifts of each RIS under the assumption $\beta_k=1, \ \forall k \in \cal K$, and then optimize the optimum binary ON-OFF vector $\cal B$ based on the obtained phase shifts. The first step, in particular, can be formulated as
\vspace{-5mm}
\begin{subequations} \label{problem1}
\begin{align}
P_{2}: & \;\;\;\;\underset{{\bf \Theta}}{\rm max}\;\; \widetilde{\gamma}_l \\
& \;\;\;\;\;\;{\rm{s}}{\rm{.t}}{\rm{.}}\;\; \beta_{k}=1, \  \forall k \in {\cal K}, \\
& \;\;\;\;\;\;\;\;\;\;\;\;  \left |e^{j \theta_{n}^k} \right | = 1, \ \forall n \in [1, N_k], \ \forall k \in {\cal K}.
 \end{align}
\end{subequations}
 
We observe that $P_{2}$ is a non-convex problem, \textcolor{black}{which can be tackled by using several methods, such as the semidefinite relaxation (SDR) method~\cite{WU01} and the successive convex approximation (SCA) method~\cite{IRS2}.} The optimal solution for the $k$th RIS is denoted by ${\Theta}_k^*$, 
and the corresponding reflection-coefficient matrix is ${\mathbf{\Phi}}^*_k$. 
With the obtained reflection-coefficient matrix, each RIS ON-OFF status is optimized via the following RIS binary control algorithm.

\subsubsection{Distributed RIS Binary Control Algorithm}
Assuming that only the $k$th RIS is ON, i.e., $\beta_k=1$, the received SINR of the signal sent from $U_l$ to the BS via the $k$th RIS is 
\begin{equation} \label{sinr_on}
\widetilde{\gamma}_l^{k}\!=\! \frac{{p_l} \left |h_{d,l}+ \textit{\textbf g}_{k} \mathbf{\Phi}_k \textit{\textbf h}_{l,k}  \right |^2 }{\sum_{m=1}^{\widetilde{\omega}_l}{p_m} \left |h_{d,m} + \xi_{m,l}^k \textit{\textbf g}_{k} \mathbf{\Phi}_k \textit{\textbf h}_{m,k}  \right |^2 + \sigma^2}.
\end{equation}

If, on the other hand, all the $K$ RISs are OFF, i.e., $\beta_k=0$ for $k \in \mathcal{K}$, the received SINR of the signal sent from $U_l$ to the BS via the direct link is
\begin{equation} \label{sinr_off}
\widetilde{\gamma}_l^{D}= \frac{{p_l} \left |h_{d,l} \right |^2 }{\sum_{m=1}^{\widetilde{\omega}_l}{p_m} \left |h_{d,m}  \right |^2 + \sigma^2}.
\end{equation}

Based on (\ref{sinr_on}) and (\ref{sinr_off}), the $k$th RIS decides whether to be ON or OFF as detailed in \textbf{Remark~\ref{R2}}.

\begin{remark} \label{R2}
The $k$th RIS should be ON if $\widetilde{\gamma}_l^{k} \geq \widetilde{\gamma}_l^{D}$ holds. This indicates, in fact, that the desired signal from $U_l$ can be enhanced via the $k$th RIS. Otherwise, the $k$th RIS should be OFF to avoid the degradation of the desired signal. In this case, the incident angle between the signals of the desired user and the interfering signals at the RIS is in general small.
\end{remark}

\renewcommand{\algorithmicrequire}{\textbf{Input:}} 
\renewcommand{\algorithmicensure}{\textbf{Output:}} 
\begin{algorithm} [thb]    
\caption{Distributed RIS Binary Control}             
\small
\label{alg1}                  
\begin{algorithmic}[1]             
\REQUIRE $\widetilde{\cal I}_l$, $\widetilde{\omega}_l$, and $\mathbf{C}_k$;
\ENSURE ${\cal B}$;\\  
\STATE Initialize $k=0$, all the RISs are ON; 
\WHILE {$k < K$}
\STATE $k\leftarrow k+1$;
\STATE Calculate ${\Theta}_k^*$ and ${\mathbf{\Phi}}^*_k$ by solving problem $P_{2}$;
\STATE Calculate $\widetilde{\gamma}_l^{k}$ and $\widetilde{\gamma}_l^{D}$ via (\ref{sinr_on}) and (\ref{sinr_off}), respectively;
\IF {$\widetilde{\gamma}_l^{k} \geq \widetilde{\gamma}_l^{D}$}
\STATE Keep the $k$th RIS ON, i.e., $\beta_k=1$;
\ELSIF {$\widetilde{\gamma}_l^{k} < \widetilde{\gamma}_l^{D}$}
\STATE Turn the $k$th RIS OFF, i.e., $\beta_k=0$;
\ENDIF
\STATE Add the $k$th RIS binary decision $\beta_k$ to $\cal B$.
\ENDWHILE    
\end{algorithmic}
\end{algorithm}

The approach for solving $P_2$ is summarized in \textbf{Algorithm~\ref{alg1}}, \textcolor{black}{which is executed at each RIS controller when the incident signal is received. Specifically, upon receiving the incident signal, each RIS controller identifies the set of interfering devices by extracting the I/Q samples from a copy of the incident signal and feeding them into the trained CNN. Based on the classification outcome of the CNN, the RIS controller can set the RIS ON-OFF state in a distributed manner.}


\subsubsection{Computational Complexity}
\textcolor{black}{The total computational complexity includes the online inference via the CNN and the iterative algorithm to solve the phase shift optimization problem $P_2$ and the RIS ON-OFF optimization problem. 
\begin{itemize}
\item Complexity for CNN online inference: The CNN is trained offline in a supervised fashion, therefore the complexity of training can be ignored. The trained CNN model has a quadratic time complexity during the inference process, i.e., ${\cal O}(M^2 C K)$, where $C$ denotes the number of layers, $M$ denotes the number of neurons, and $K$ denotes the total number of RISs.
\item Complexity for solving the phase shift optimization problem $P_2$: To solve the problem $P_2$, the complexity lies in computing the optimal phase shift at each iteration of the optimization method, e.g., the SCA method~\cite{IRS2} whose complexity is ${\cal O}(Qz)$, where $Q=\sum_{k=1}^{K}N_k$ denotes the total number of elements of all the RISs, and $z$ is the total number of the iterations required.
\item Complexity for solving the RIS ON-OFF optimization problem: 
 Since $\widetilde{\gamma}_l^{k}$ and $\widetilde{\gamma}_l^{D}$ need to be calculated via (\ref{sinr_on}) and (\ref{sinr_off}), respectively, the computational complexity of solving the RIS ON-OFF optimization problem required at each RIS controller is ${\cal O}(K)$.
\end{itemize}}

\textcolor{black}{As a result, the total complexity is ${\cal O}(M^2 C K+Qz+K)$, which grows linearly with the total number of RISs.}

\section{Simulation Results}\label{results}
In this section, we first evaluate the inference accuracy of the trained CNN model \textcolor{black}{and the computational complexity of the proposed ISL-based DRBC algorithm.} Then we validate the benefits of deploying the ISL-based DRBC algorithm.
\vspace{-2mm}
\subsection{CNN Testing Results}
We trained the CNN with the $80\%$ of collected RF data set which contains about $800$ million I and Q samples (training set), validated it by using $10\%$ of the dataset (validation set), and tested it by using $10\%$ of the dataset (testing set) each corresponding to about $100$ million of the I and Q samples. \textcolor{black}{The trained CNN model consists of two convolutional (Conv) layers with ReLU activation functions, followed by two dense fully connected (FC) layers. In particular, the trained CNN model contain $256$ filters (1$\times$3) in the first Conv layer, $128$ filters (1$\times$3) in the second Conv layer, $256$ neurons in the first FC layer, and $9$ neurons in the second FC layer (output).}


The classification accuracy of the trained CNN is analyzed in Table~\ref{Inference_accuracy}, by considering a two-user scenario. The window size (i.e., the number of time steps of the collected RF data) is $32$, $128$, and $512$, respectively. We observe from Table~\ref{Inference_accuracy} that the online inference accuracy is, in general, greater than $95\%$ in the  considered scenario.  \textcolor{black}{Compared to other classes, the `Idle' class has the main characteristic that no user transmits and only background noise exists. Due to the {distinguishable pattern compared to the other three classes}, the CNN model predicts the `Idle' class perfectly.}
 \begin{table}[t]\footnotesize
  \captionsetup{font={small}} 
\caption{Inference accuracy of the trained CNN model.}   
 \label{Inference_accuracy}
\centering
\footnotesize
\begin{tabular}{|l | l | l | l| } 
\hline
\textbf{Scenarios} & $w=32$ & $w=128$ & $w=512$\\
\hline
Idle & 100.00$\%$ & 100.00$\%$ & 100.00$\%$ \\
\hline 
Only $U_1$   &  98.35$\%$ & 98.04$\%$ & 96.21$\%$\\
\hline
Only $U_2$    & 96.09$\%$ & 96.12$\%$ & 95.64$\%$ \\
\hline
$U_1\!+\!U_2$ & 99.66$\%$ & 99.76$\%$ & 99.93$\%$  \\
\hline
\end{tabular}
\end{table}
\vspace{-4mm}
  \subsection{\textcolor{black}{Computation Time}}
\textcolor{black}{The proposed ISL-based DRBC algorithm allows us to obtain the optimal ON-OFF status of each RIS. As detailed in previous text, the optimal ON-OFF status of the RISs can be formulated as the solution of a non-convex MINLP, which is usually challenging to solve~\cite{IRS2}.
In this section, we compare the proposed ISL-based DRBC algorithm against the spatial branch and bound (sBB) method, which is often employed to solve non-convex MINLP~\cite{MINLP}, in terms of computation time.}


\textcolor{black}{The comparison of the average computation time (defined as $t=\frac{\rm Total \ time \ consumption}{\rm Total \ number \ of \ computations}$) between the proposed ISL-based DRBC algorithm and the traditional sBB method is conducted on the same hardware platform that consists of an Intel Xeon(R) CPU E5-2650@2.0 GHz x 16. The obtained results are illustrated in Table~\ref{ac1}. Compared to the traditional sBB method, the ISL-based DRBC algorithm results in much lower computation time while still yielding the optimal ON-OFF status for each RIS. The computation time of the proposed algorithm is less than one-thousandth of the computation time of the sBB method when the number of RISs varies from $2$ to $5$.}

\begin{table}[thb]  
\renewcommand{\arraystretch}{1.2}
  \captionsetup{font={small }} 
\caption{ \textcolor{black}{Computation time ($ms$) of the proposed ISL-based DRBC algorithm and the traditional sBB method }} 
\label{ac1}  
\centering  
\footnotesize
\begin{tabular}{|c|c|c|c|}
\hline
\diagbox{$K$}{$t$} & Traditional sBB method & Proposed DRBC algorithm\\
 \hline
2 & $ 14.1 $ & $ 2.60 \times 10^{-3}$\\
\hline
3  & $ 14.2 $  & $2.55 \times 10^{-3} $ \\
\hline
4  & $ 14.5 $ & $2.40 \times 10^{-3} $  \\
\hline
5 & $ 15.2$ & $ 5.15 \times 10^{-3}$\\
\hline
\end{tabular}  
\end{table} 
\vspace{-8mm}
\subsection{Performance Evaluation}
\subsubsection{Simulation Setting} 
The simulation model consists of $K$ RISs, one desired user ($U_1$), and one interfering user ($U_2$). Each RIS consists of $256$ elements and all the $K$ RISs are 
equally spaced by $5$ m in vertical direction. The distances from BS and $U_1$ to the RIS center are $80$ m and $60$ m, respectively. The incident angle between the BS and $U_1$ at the RIS is $150\rm{^o}$, and the incident angle between $U_2$ and $U_1$ is $\lambda \in [0, 150\rm{^o}]$. We assume that $\xi_{m,l}^k$ is linearly inversely proportional to $\lambda$. The channel parameters are selected according to the 3GPP Urban Micro standard~\cite{3GPP}, which describes the path loss for both line-of-sight and non-line-of-sight components~\cite{tvt_zhang}. The transmission power of $U_1$ is $20$ dBm, the noise power $\sigma^2$ is -$94$ dBm, the carrier frequency is $3$ GHz, and the reflection amplitude is equal to one.

\subsubsection{Simulation Results}
\begin{figure}[t]
\centering
  \captionsetup{font={footnotesize }}
\subfigure[]{
\includegraphics[width=2.35in,height=1.82in]{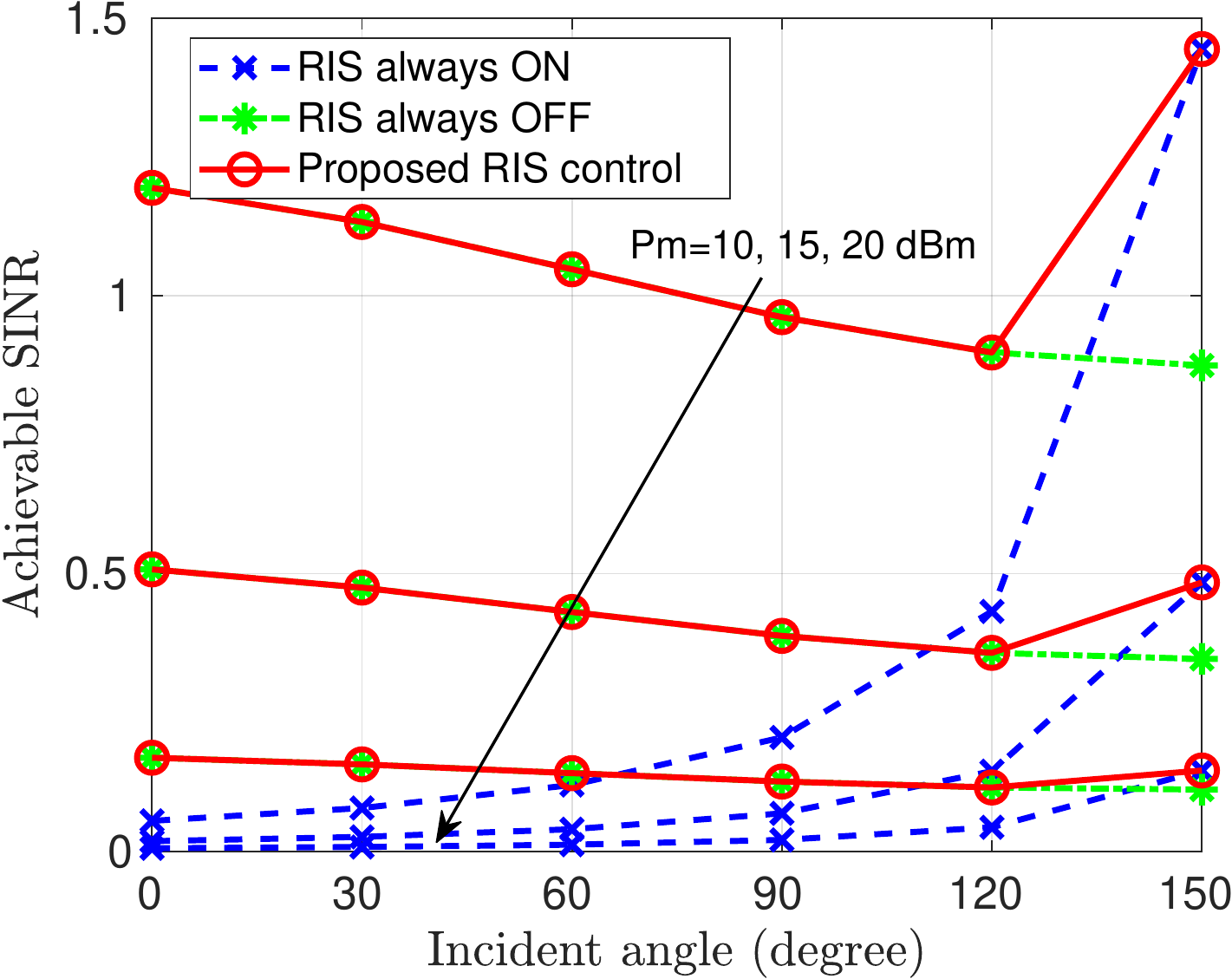}}
\hspace{2.85mm}
\subfigure[]{
\includegraphics[width=2.35in,height=1.82in]{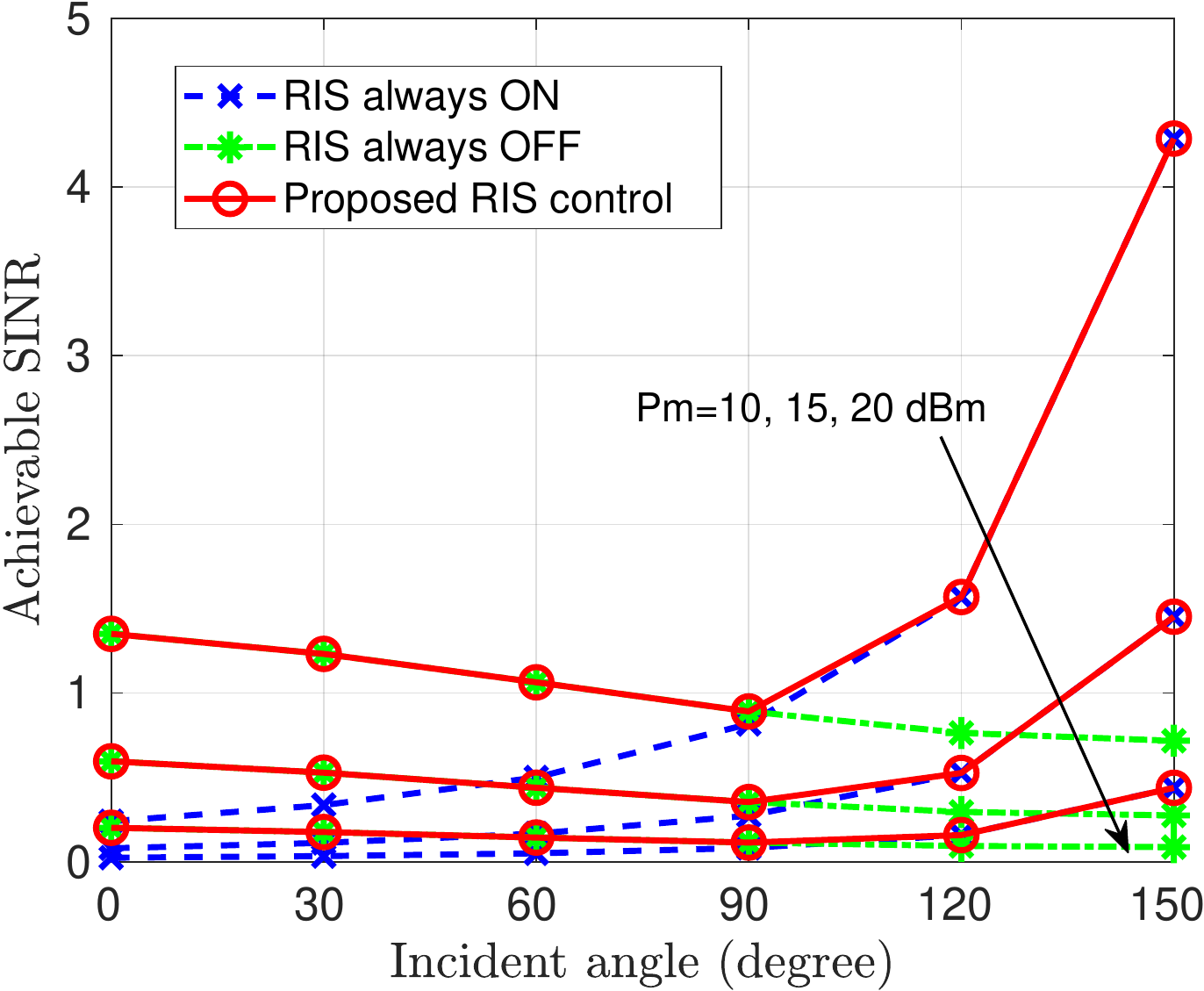}}
\vspace{-0.05in}
\caption{Achievable SINR vs. $\lambda$ for $K=1$. The distance between the interfering user ($U_2$) and the RIS is $10$ m in (a), and $5$ m in (b), respectively.} 
\label{expriment}
\end{figure}

We evaluate the performance of the proposed ISL-aided algorithm by comparing it with two benchmarks: `RIS always ON' and `RIS always OFF'. Figs.~\ref{expriment}(a)-(b) depict the achievable SINR versus the angle of incidence ($\lambda$) for $K=1$ and $p_m=10, 15, 20$ dBm. 
We observe that the SINR first gradually decreases as $\lambda$ increases due to the reduction of the distance between $U_2$ and BS, and then increases due to the perfect interference elimination at the RIS. When $\lambda$ is small in particular, 
the RIS is prone to be OFF since the interference reflected via the RIS is more pronounced. If $\lambda$ is large, on the other hand, the impact of the interference is reduced and it is more probable that the RIS is ON. \textcolor{black}{In general terms, however, the impact of $\lambda$ on the system performance is still an open issue, whose analysis is postponed to a future research work.}

\begin{figure}[t]
\centering
  \captionsetup{font={footnotesize }}
\subfigure[]{
\hspace{-0.1in}
\includegraphics[width=2.05in,height=1.62in]{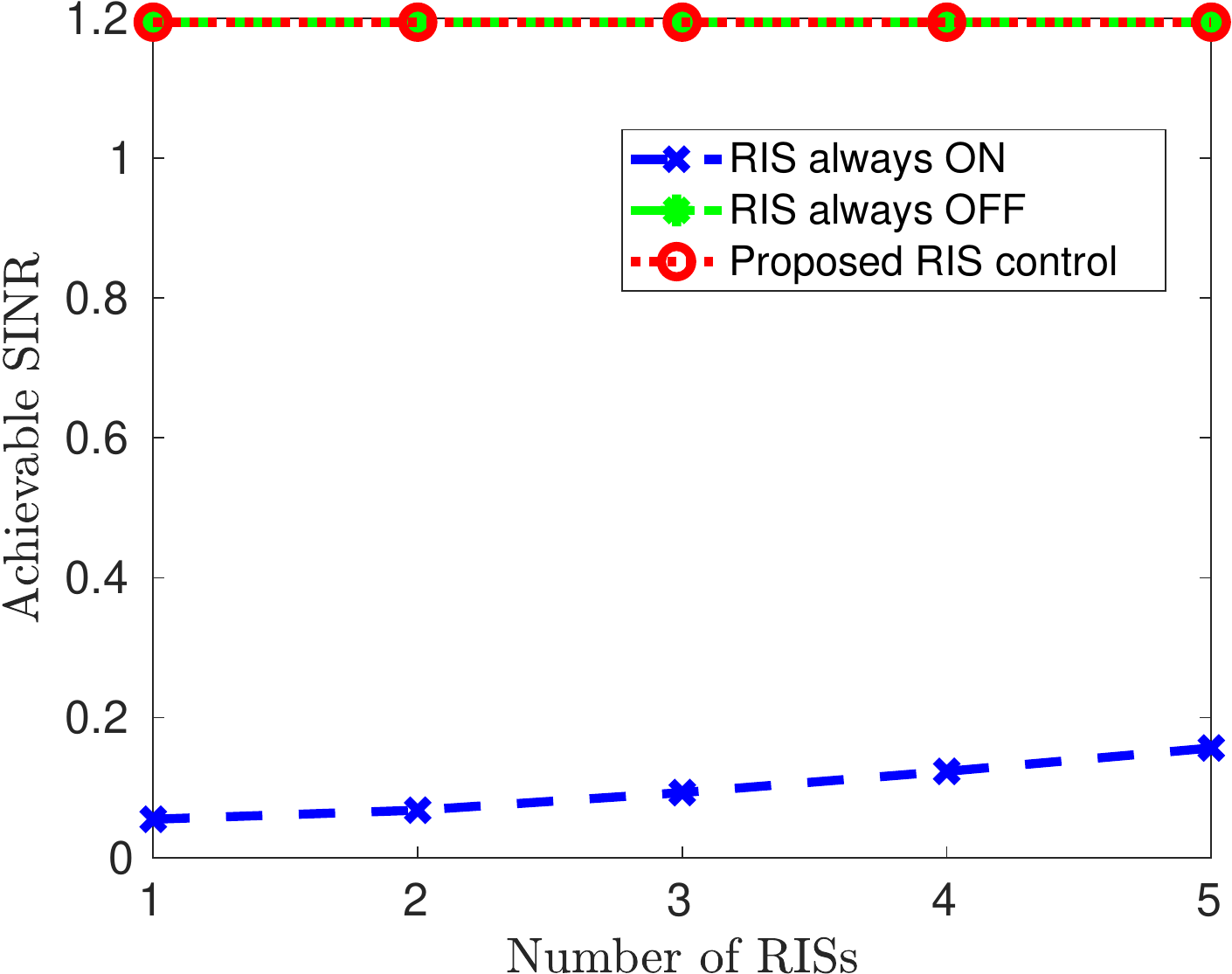}}
\hspace{-0.1in}
\subfigure[]{
\includegraphics[width=2.05in,height=1.62in]{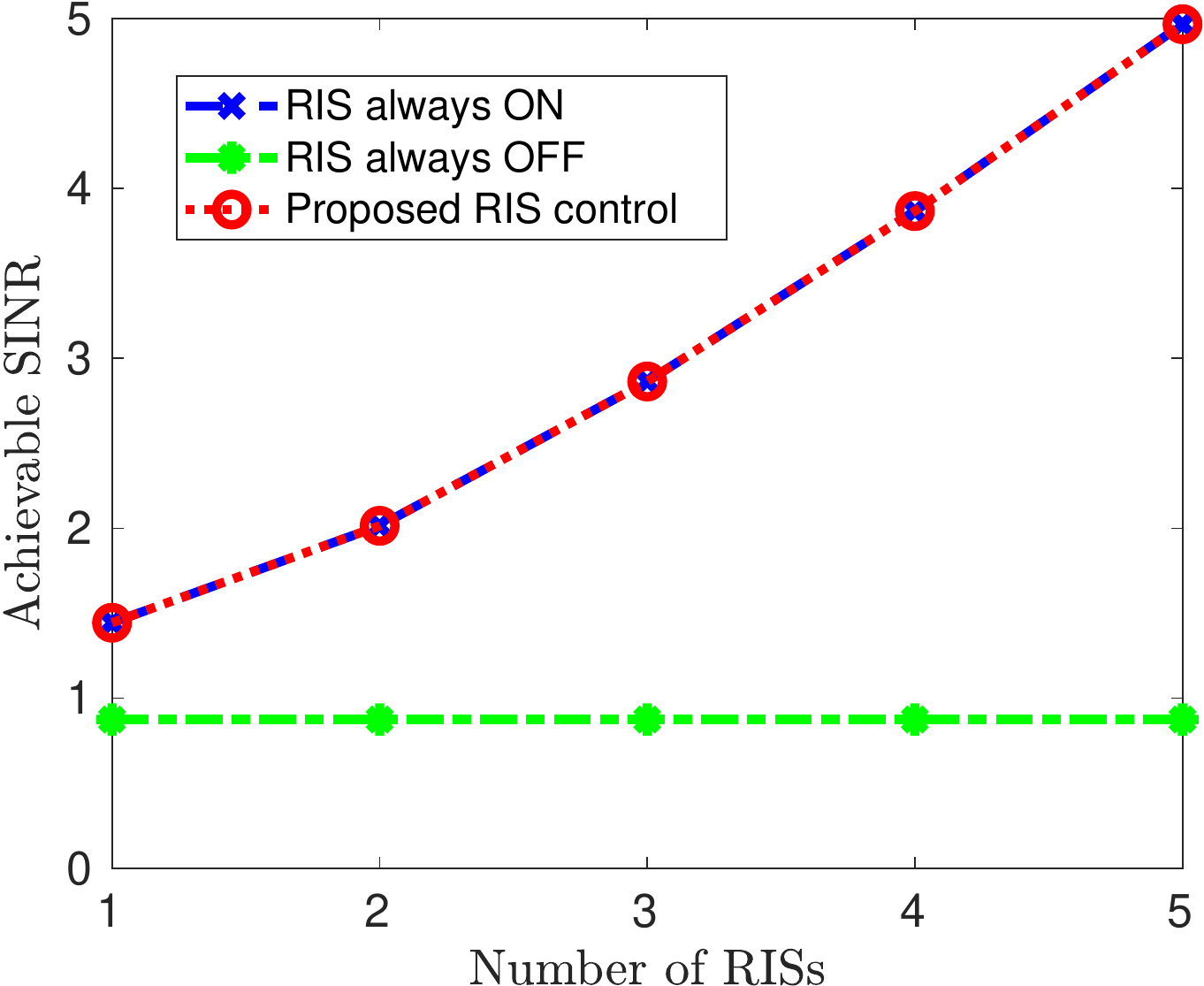}}
\hspace{-0.1in}
\subfigure[]{
\includegraphics[width=2.05in,height=1.62in]{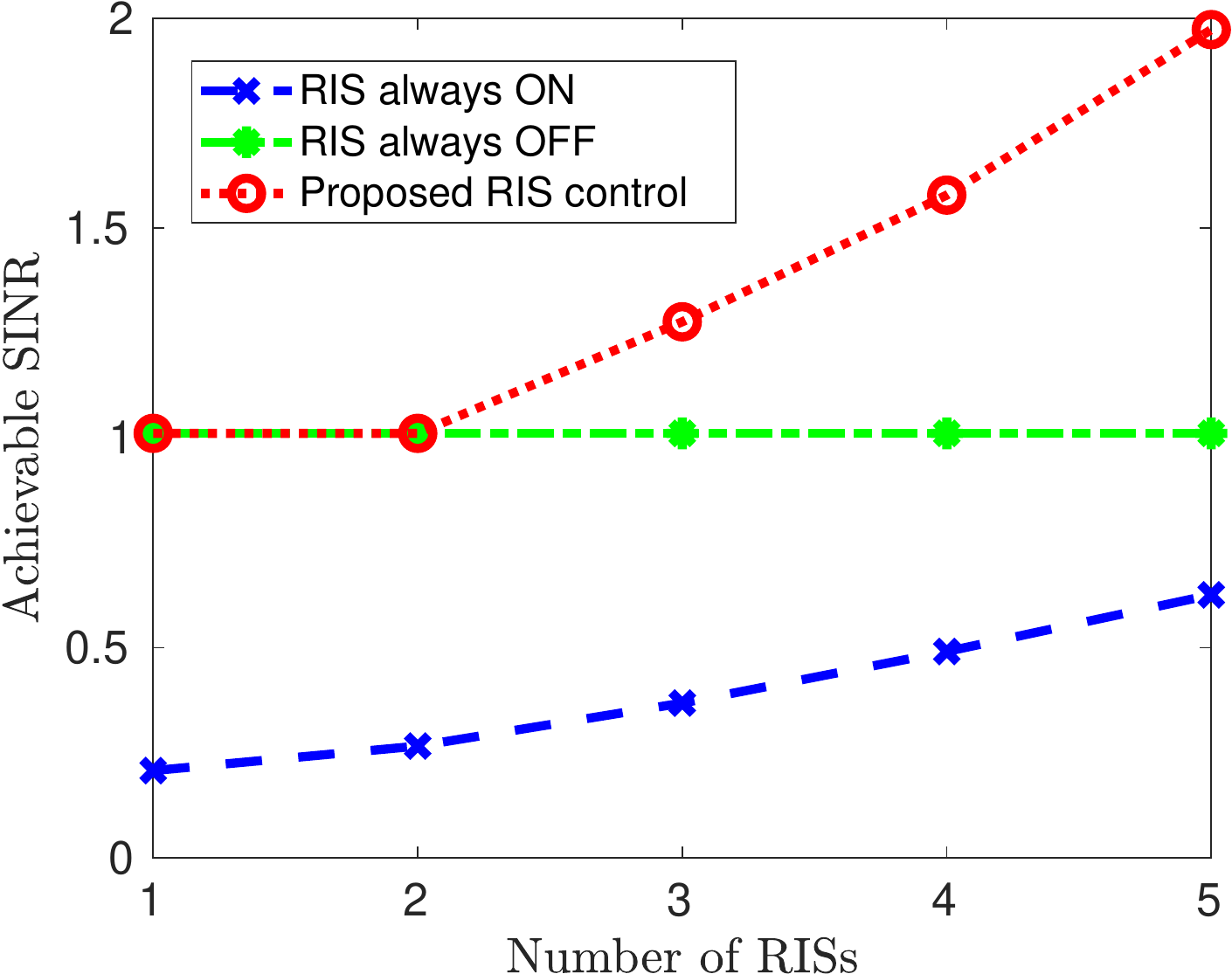}}
\vspace{-0.05in}
\caption{SINR vs. $K$ for $p_m=10$ dBm. $\lambda=0$ is shown in (a), $\lambda=150\rm{^o}$ is shown in (b), and $\lambda$ randomly selected in $[30\rm{^o}, 120\rm{^o}]$ 
is shown in (c).} 
\label{expriment1}
\end{figure}

In Figs.~\ref{expriment1}(a)-(c), the SINR versus $K$ is illustrated, where $p_m=10$ dBm, the distance between $U_2$ and the RIS is $5$ m. We observe that the RISs are always OFF if $\lambda=0$, since the impact of the interference is too high. If $\lambda$ is very large, e.g., $\lambda=150\rm{^o}$ in Fig.~\ref{expriment1}(b), the impact of the interference is low and the RISs are always ON. 
When $\lambda$ is randomly selected, e.g., $\lambda \in [30\rm{^o}, 120\rm{^o}]$ in Fig.~\ref{expriment1}(c), the SINR obtained by the ISL algorithm increases with $K$ and outperforms the two benchmarks, by about $100\%$ with respect to `RIS always OFF' and by nearly $300\%$ with respect to `RIS always ON' when $K=5$.  
From Figs~\ref{expriment1}(a)-(c), we conclude that the performance of the proposed ISL algorithm largely depends on $\lambda$, which impacts the interference cancellation at the RISs.

\vspace{-0.1in}
 
\section{Conclusion \textcolor{black}{and Future Work}}
\label{conclusion}

In this paper, we introduced an ISL algorithm that uses appropriately trained CNNs for the interference management in RIS-aided multi-user uplink networks. With the aid of the ISL algorithm, the RISs are capable of inferring the interfering signals directly from the incident signals. A distributed control algorithm was proposed to maximize the received SINR by dynamically configuring the binary status of the RIS elements. Simulation results validated the performance improvement offered by the proposed ISL-aided RIS approach. 
\textcolor{black}{\textcolor{black}{Offline training is only a candidate way to train a CNN, which may need to be retrained when RF data distribution changes significantly. This issue may be avoided by using online training methods, such as federated learning, which is a promising method for application in dynamical wireless environments. } }


\end{document}